\documentclass[a4paper, amsfonts, amssymb, amsmath, reprint, showkeys, nofootinbib, twoside]{revtex4-1}
\usepackage{graphicx}  
\usepackage{dcolumn}   
\usepackage{bm}        
\usepackage{dutchcal}
\usepackage[toc,page]{appendix}
\usepackage{braket}
\usepackage{mathtools}

\begin{document}

\title{Quantization of resistivity as consequence of symmetry invariance }
\author{Jorge A. Lizarraga}
\email{jorge\_lizarraga@icf.unam.mx\\}
\affiliation{Instituto de Ciencias F\'isicas, Universidad Nacional Aut\'onoma de M\'exico, Cuernavaca, M\'exico}

\date{\today}

\begin{abstract}
The Schrödinger equation for an electron under the influence of an electromagnetic field is analyzed based on the conserved operators of the system when the magnetic field is described by Landau's gauge. It is shown that the Lorentz force can be recovered only if two conserved generalized momentum operators are considered: one along the $x$-axis and the second one along $y$-axis; otherwise, the system cannot be fully described. Based on the general solution found, a ground state is built which has the characteristic of having quantized resistivity proportional to integer multiples of the von Klitzing's constant when it is invariant under a unitary transform.
\end{abstract}
\maketitle

\section{Introduction}

Wave functions for the Schrödinger equation involving a constant electromagnetic field have been obtained using different methods such as perturbation theory \cite{messiah1999,sakurai2017}. However, the most common way to solve this equation is by using Landau's ansatz, which is based on the search for conserved operators for the specific choice of gauge and electric potential, leading to a separable variable solution of plane waves along the $x$-axis (or $y$-axis) when the magnetic field is described using Landau's gauge, and plane waves along the angular coordinate when described with the symmetric gauge \cite{landau1930,lifshitz1981,ciftja2020detailed}. This last case is the one we are interested in analyzing. In this work, the magnetic field is described using Landau's gauge, and the electric field is selected in such a way that the momentum ${\hat p}_{x}$ is conserved, making Landau's ansatz still applicable. However, based on the argument that the force acting upon the particle must be described by Lorentz's law, it turns out that a second conserved operator is needed. This operator has the characteristic of being time-dependent, leading to an interesting situation where the wave function found is time-dependent, i.e., the spatial and time coordinates are mixed in such a way that they cannot be written as the product of functions which depends on each variable respectively.
The conserved operators of the system are also useful for analyzing its degeneracy, as they act as generators of solutions. They can be used to construct the general solution for the problem, which can later be simplified to find a time-dependent ground state of the system. Additionally, they define a set of unitary operators that characterize the symmetries of the system. This state exhibits the characteristic of having quantized resistivity when it is invariant under a unitary transformation given by the time-dependent conserved operator presented. This property is reminiscent of the experimental resistivity reported by Störmer \cite{tsui1982two}. It's also important to note that the fact that this ground state can be found due to the degeneracy of the system matches Tao and Wu's analysis, where the conductance is accurately quantized to a rational value \cite{tao1984gauge}.

This work is organized as follows: In Section II, the conserved quantities of the Hamiltonian are analyzed, and it is shown that Lorentz's Force can only be described with two conserved quantities. Section III demonstrates how to obtain solutions for the system using the conserved operators, including Landau's solution and a non-separable variable solution. Section IV utilizes the conserved operators to analyze the degeneracy of the system and construct the general solution for this problem. Finally, in Section V, a ground state is constructed such that it exhibits a quantized resistivity proportional to integer multiples of the von Klitzing's constant \cite{klitzing1980new}.

\section{Hamiltonian and conserved quantities}

The Hamiltonian we are interested in is written in CGS units as:
\begin{equation}\label{H}
{\hat H}=\frac{1}{2m}\left({\bf \hat P}-\frac{q}{c}{\bf A}\right)^2+V,
\end{equation}
where $m$ is the mass of the particle, $q$ is the charge of the particle, $c$ is the speed of light, ${\bf\hat P}=({\hat p}_{x},{\hat p}_{y},{\hat p}_{z})=-i\hbar\nabla$ is the momentum operator, ${\bf A}=(A_{x},A_{y},A_{z})$ is the vector potential (or gauge) such that the magnetic field is given by it curl, that is ${\bf B}=\nabla\times{\bf A}$ and $V=-q{\cal E}y$ is the potential energy. We set our magnetic field such that is parallel to the $z$ axis, ${\bf B}=B{\hat k}$, where $B$ is constant and described by the Landau's gauge ${\bf A}=B(-y,0,0)$. From now on we are only interested in the dynamics of the particle in the $x-y$ plane, therefore we set $p_{z}=0$ and $A_{z}=0$. Hence, the Hamiltonian can be written as:
\begin{equation}\label{H2}
{\bf\hat H}=\frac{1}{2m}\left(\left({\hat p}_{x}+m\omega_{c}y\right)^2+{\hat p}_{y}^{2}\right)-q{\cal E}y,
\end{equation}\\
where the cyclotron frequency $\omega_{c}=qB/mc$ is defined. Then the following operators can be defined
\begin{equation}\label{px}
{\hat\pi'}_{x}=\hat{p}_x,
\end{equation}
\begin{equation}\label{py}
{\hat\pi}'_{y}=\hat{p}_y+m\omega_{c}x-q{\cal E}t,
\end{equation}
and
\begin{equation}\label{E}
{\hat E}=i\hbar\frac{\partial}{\partial t},
\end{equation}

\noindent such that all of them are conserved, that is
\begin{equation}\label{consv}
\frac{d{\hat\pi}'_{x}}{dt}=0,\quad\frac{d{\hat\pi}'_{y}}{dt}=0\quad \text{and}\quad \frac{d{\hat E}}{dt}=0.
\end{equation}

\noindent In the literature, it is typically only considered the operator (\ref{px}) for the search of the basis of the system (see reference \cite{lifshitz1981} and all the works based on it). However, the operator (\ref{py}) is also necessary for a full description of the system. This assertion can be demonstrated by calculating the total variation of the coordinates.
\begin{equation}\label{velox}
\frac{d x}{dt}=\frac{1}{m}\left(\hat{p}_x+m\omega_{c}y\right),
\end{equation}\\
and
\begin{equation}\label{veloy}
\frac{d y}{dt}=\frac{1}{m}\hat{p}_y,
\end{equation}\\
then, from the first conservation equality in Eq. (\ref{consv}) and equality Eq. (\ref{velox}), one can get the next expression
\begin{equation}\label{}
m\frac{d^{2} x}{dt^{2}}=m\omega_{c}\frac{dy}{dt},
\end{equation}\\
and using the second conservation equation in Eq. (\ref{consv}) and  Eq. (\ref{veloy}), one obtains
\begin{equation}\label{}
m\frac{d^{2} y}{dt^{2}}=-m\omega_{c}\frac{dx}{dt}+q{\cal E}.
\end{equation}

\noindent Now, the last two equations can be combined into a vectorial form that can be rewritten as Newton's second law
\begin{equation}\label{}
{\bf F}=\frac{q}{c}{\bf v}\times{\bf B}+q{\bf E}.
\end{equation}
And Lorentz's force is obtained. Hence, both conserved operators (\ref{px}) and (\ref{py}) are equally important to fully describe the system.

\section{Solutions of the system}

Once the conserved operators are known, we can use them to define a set of eigenvalue equations that can be used to find the basis of the system. For simplicity, we write them as:
\begin{equation}\label{eigen_eq_1}
{\hat\pi'_{x}}\psi=-m\omega_{c}\delta y\psi,
\end{equation}

\begin{equation}\label{eigen_eq_2}
{\hat\pi'_{y}}\overline\psi=m\omega_{c}\delta x\overline\psi,
\end{equation}

\noindent where $\delta x$ and $\delta y$ are real constants. Solving Eq. (\ref{eigen_eq_1}), one find that 
\begin{equation}\label{}
\psi(x,y)={\cal C}(y)e^{-i\frac{m\omega_{c}}{\hbar}x\delta y},
\end{equation}
where the function ${\cal C}$ can be determined by substituting it in the eigenvalue equation ${\bf\hat H}\psi=E\psi$ where the Hamiltonian (\ref{H2}) is used. The resulting equation that satisfies this function is the displaced harmonic oscillator as is already known \cite{lifshitz1981}, therefore, defining the function
\begin{equation}\label{harm_oscil_func}
\varphi_{n}(\xi)=\frac{1}{\sqrt{2^{n}n!}}\left(\frac{m\omega}{\pi\hbar}\right)^{1/4}\exp\left(-\frac{\xi^{2}}{2}\right)H_{n}(\xi),
\end{equation}\\
where $n\in\mathbb{N}$ and $H_{n}$ are the Hermite polynomials. Then 
\begin{equation}\label{}
{\cal C}(y)=\varphi_{n}\left(\sqrt{\frac{m\omega_{c}}{\hbar}}\left(y-\delta y-\frac{q{\cal E}}{m\omega_{c}^{2}}\right)\right),
\end{equation}\\
being the eigenvalues
\begin{equation}\label{}
E_{n}=\hbar\omega_{c}\left(n+\frac{1}{2}\right)-\frac{1}{2m}\frac{q^{2}{\cal E}^{2}}{\omega_{c}^{2}}-q{\cal E}\delta y.
\end{equation}\\

\noindent On the other hand, solving Eq. (\ref{eigen_eq_2})
\begin{equation}\label{psibar}
\overline\psi=\overline{\cal C}(x,t)\exp\left(-i\frac{m\omega_{c}}{\hbar}(x-\delta x)y+i\frac{q{\cal E}}{\hbar}ty\right),
\end{equation}
note that in this case, the wavefunction is time-dependent and so is the function 
 $\overline{\cal C}$. Therefore, to determine this function, it is necessary to substitute the above wavefunction into the complete Schrödinger equation, ${\bf\hat H}\overline\psi={\hat E}\overline\psi$, instead of the stationary one. By doing so, the following equation is obtained:
\begin{equation}\label{}
\frac{1}{2m}\left({\hat p}_{x}^{2}\overline{\cal C}+(m\omega_{c}(x-\delta x)-q{\cal E}t)^{2}\overline{\cal C}\right)={\hat E}\overline{\cal C}.
\end{equation}
Even though in the above equation the variables $x$ and $t$ are not longer separable, the function $\overline{C}$ still has the information of the energies of the system (denoted by $E'$), therefore, is helpful to rewrite the function as 
\begin{equation}\label{}
\overline{\cal C}(x,t)=\exp\left(-i\frac{E'}{\hbar}t\right)D(x,t).
\end{equation}
Substituting this expression in the last equation and defining the variable 
\begin{equation}\label{}
\xi=\sqrt{\frac{m\omega_{c}}{\hbar}}\left(x-\delta x-\frac{q{\cal E}}{m\omega_{c}}t\right),
\end{equation}
an equations for $D$ is obtained
\begin{equation}\label{}
-\frac{\partial^{2}D}{\partial\xi^{2}}+\xi^{2}D=-i2\frac{q{\cal E}}{m\omega_{c}^{2}}\sqrt{\frac{m\omega_{c}}{\hbar}}\frac{\partial D}{\partial\xi}+\frac{2E'}{\hbar\omega_{c}}D.
\end{equation}
Applying Fourier transform, defined as
\begin{equation}\label{}
{\cal F}\{f\}=\frac{1}{\sqrt{2\pi}}\int\limits_{\mathbb{R}}e^{ik\xi}f(\xi) d\xi,
\end{equation}
to this equation, defining $\overline D={\cal F}\{D\}$ and the constant 
\begin{equation}\label{}
a=\frac{q{\cal E}}{m\omega_{c}^{2}}\sqrt{\frac{m\omega_{c}}{\hbar}},
\end{equation}
one can write down the displaced harmonic oscillator equation
\begin{equation}\label{}
-\frac{\partial^{2}\overline D}{\partial k^{2}}+(k+a)^{2}\overline D=\left(\frac{2 E'}{\hbar\omega_{c}}+a^{2}\right)\overline D,
\end{equation}
which its solutions are given again by Eq. (\ref{harm_oscil_func})
\begin{equation}\label{D_linea}
\overline D=\varphi_{n}(k+a),
\end{equation}
having the eigenvalues 
\begin{equation}\label{}
E_{n}'=\hbar\omega_{c}\left(n+\frac{1}{2}\right)-\frac{1}{2m}\frac{q^{2}{\cal E}^{2}}{\omega_{c}^{2}}.
\end{equation}
However, there is one more step we need to take in order to find the solution in the original variables. We can do so by performing the inverse Fourier transform of the expression in Eq. (\ref{D_linea}). Knowing that the inverse Fourier transform of a harmonic oscillator is another harmonic oscillator, we can write: 
\begin{equation}\label{D}
D=e^{ia\xi}\varphi_{n}(\xi).
\end{equation}

\section{Symmetries and degeneracy}
Once the process to obtain the solutions for this system from its conserved properties has been illustrated, it is helpful to gather the results in order to compare them. From conserved operator (\ref{px}), one can find the following wave functions
\begin{equation}\label{wf1}
\begin{aligned}
\psi_{n}=&\exp\left({-i\frac{E_{n}}{\hbar}t-i\frac{m\omega_{c}}{\hbar}x\delta y}\right)\times\\
&\varphi_{n}\left(\sqrt{\frac{m\omega_{c}}{\hbar}}\left(y-\delta y-\frac{q{\cal E}}{m\omega_{c}^{2}}\right)\right),
\end{aligned}
\end{equation}
having the energies defined as 
\begin{equation}\label{Energy1}
E_{n}=\hbar\omega_{c}\left(n+\frac{1}{2}\right)-\frac{1}{2m}\frac{q^{2}{\cal E}^{2}}{\omega_{c}^{2}}-q{\cal E}\delta y.
\end{equation}\\
On the other hand, from conserved operator (\ref{py}), one obtain the following wave functions 

\begin{equation}\label{wf2}
\begin{aligned}
\overline\psi_{n} =&\exp\bigg(-i\frac{E_{n}'}{\hbar}t-i\frac{m\omega_{c}}{\hbar}(x-\delta x)y+i\frac{q{\cal E}}{\hbar}ty\bigg)\times\\
&\exp\bigg(i\frac{q{\cal E}}{\hbar\omega_{c}}\left(x-\delta x-\frac{q{\cal E}}{m\omega_{c}}t\right)\bigg)\times\\
&\varphi_{n}\left(\sqrt{\frac{m\omega_{c}}{\hbar}}\left(x-\delta x-\frac{q{\cal E}}{m\omega_{c}}t\right)\right),
\end{aligned}
\end{equation}

\noindent where the energies were defined as
\begin{equation}\label{Energy2}
E_{n}'=\hbar\omega_{c}\left(n+\frac{1}{2}\right)-\frac{1}{2m}\frac{q^{2}{\cal E}^{2}}{\omega_{c}^{2}}.
\end{equation}
The first difference to note is that Eq. (\ref{wf1}) and Eq. (\ref{wf2}) involve oscillations along the $x$-axis and $y$-axis respectively, this matches with the classical idea that the particle is describing a drifted circle. Hence, each wave function can be seen as kind of projection of this motion along its respective axis. However, there is another notable difference, the energies of the system are not exactly the same. Comparing expressions in Eq. (\ref{Energy1}) and Eq. (\ref{Energy2}), one can realize that $E_{n}'-E_{n}=q{\cal E}\delta y$. This insight suggests that there must be an even more fundamental form of the solutions. The conserved operators in Eq. (\ref{px}), Eq. (\ref{py}), and Eq. (\ref{E}) can be used to define a set of unitary operators.
\begin{equation}\label{unitary1}
{\hat U}_{x}=\exp\left(-i\frac{\delta x}{\hbar}{\hat\pi}_{x}\right),
\end{equation}

\begin{equation}\label{unitary2}
{\hat U}_{y}=\exp\left(-i\frac{\delta y}{\hbar}{\hat\pi}_{y}\right),
\end{equation}
and
\begin{equation}\label{unitary3}
{\hat U}_{t}=\exp\left(i\frac{\delta t}{\hbar}{\hat E}\right).
\end{equation}
All of the above unitary operators define the symmetries of the system in the Heisenberg picture, meaning that the Schr\"odinger equation remains invariant under a transformation of the form:
\begin{equation}\label{}
{\hat H}-{\hat E}={\hat U}_{i}^{\dagger}({\hat H}-{\hat E}){\hat U}_{i},
\end{equation}
such that $i=x,y,t$. However, in Schr\"odinger picture, where the unitary operators are applied to the wave function, one can define $\zeta_{n}={\hat U}_{y}^{\dagger}\psi_{n}$ being
\begin{equation}\label{zeta1}
\zeta_{n}=\exp\left(-i\frac{E_{n}'}{\hbar}t\right)\varphi_{n}\left(\sqrt{\frac{m\omega_{c}}{\hbar}}\left(y-\frac{q{\cal E}}{m\omega_{c}^{2}}\right)\right).
\end{equation}
Similarly, we can define $\overline\zeta_{n}={\hat U}_{x}^{\dagger}\overline\psi_{n}$ and write down the function 
\begin{equation}\label{zeta2}
\begin{aligned}
\overline\zeta_{n}=&\exp\left(-i\frac{E_{n}'}{\hbar}t-i\frac{m\omega_{c}}{\hbar}xy+i\frac{q{\cal E}}{\hbar}ty+i\frac{q{\cal E}}{\hbar\omega_{c}}\left(x-\frac{q{\cal E}}{m\omega_{c}}t\right)\right)\\
&\varphi_{n}\left(\sqrt{\frac{m\omega_{c}}{\hbar}}\left(x-\frac{q{\cal E}}{m\omega_{c}}t\right)\right).
\end{aligned}
\end{equation}

\noindent Both wave functions in Eq. (\ref{zeta1}) and Eq. (\ref{zeta2}) are the fundamental solutions of the Schr\"odinger equation, defined by the Hamiltonian in Eq. (\ref{H2}), and they have the same energies given by Eq. (\ref{Energy2}).

\noindent This situation arises from the fact that the parameter $\delta y$ causes the states to be continuously degenerated. However, there is another discrete degeneracy in this system, which we will now analyze.

\noindent When it comes to time-independent operators, like Eq. (\ref{px}), it is said to be conserved for one out of two reasons: 1) because it shares a basis with the Hamiltonian, or 2) because it is an eigenfunction generator. On the other hand, the situation with time-dependent conserved operators is analogous, but instead of being eigenfunction generators, they are generators of solutions of the time-dependent Schrödinger equation. This characteristic arises from the fact that the wave function cannot share basis with all the operators at the same time.

\noindent Therefore, for the wave function in Eq. (\ref{zeta1}), the generator of solutions is the operator in Eq. (\ref{py}), i.e. ${\bf\hat H}({\hat\pi}_{y}\zeta_{n})={\hat E}({\hat\pi}_{y}\zeta_{n})$. This can be easily generalized to any $j\in\mathbb{N}$ applications of the operator, that is
\begin{equation}\label{}
{\bf\hat H}({\hat\pi}_{y}^{j}\zeta_{n})={\hat E}({\hat\pi}_{y}^{j}\zeta_{n}).
\end{equation}
On the other hand, for the wave equation Eq. (\ref{zeta2}) the generator of solutions is the operator in Eq. (\ref{px}) that is 
\begin{equation}\label{}
{\bf\hat H}({\hat\pi}_{x}^{j}\overline\zeta_{n})={\hat E}({\hat\pi}_{x}^{j}\overline\zeta_{n}).
\end{equation}
The same situation happens with the operator Eq. (\ref{E}), when we apply it to a time-dependent wave function. While the application of Eq. (\ref{E}) to the wave function Eq. (\ref{zeta1}) is elementary, it is not when it is applied to the wave function ${\hat\pi}_{y}^{j}\zeta_{n}$, since the result is not proportional to the original function. Hence, another set of solutions of the Schr\"odinger equation is given by the application of the conserved operator Eq. (\ref{E}) as follows 
\begin{equation}\label{}
{\bf\hat H}({\hat E}^{j'}{\hat\pi}_{y}^{j}\zeta_{n})={\hat E}({\hat E}^{j'}{\hat\pi}_{y}^{j}\zeta_{n}).
\end{equation}
such that $j'\in\mathbb{N}$. Similarly, applying Eq. (\ref{E}) to the functions ${\hat\pi}_{y}^{j}\overline\zeta_{n}$, due to the conservation of the operator, we have that 
\begin{equation}\label{}
{\bf\hat H}({\hat E}^{j'}{\hat\pi}_{y}^{j}\overline\zeta_{n})={\hat E}({\hat E}^{j'}{\hat\pi}_{y}^{j}\overline\zeta_{n}).
\end{equation}

\noindent Finally, the general solution for this system can be written as
\begin{equation}\label{Gen_sol}
\Psi=\sum_{n,j,j'}c_{n,j,j'}{\hat E}^{j'}{\hat\pi}_{y}^{j}\zeta_{n}+\sum_{n,j,j'}\overline{c}_{n,j,j'}{\hat E}^{j'}{\hat\pi}_{x}^{j}\overline\zeta_{n},
\end{equation}
where $c_{n,j,j'}$ and $\overline c_{n,j,j'}$ are constants.

\section{Quantization of resistivity}

Once we have our general solution, Eq. (\ref{Gen_sol}), we can use it to calculate the electric current produced by it, defined as 
\begin{equation}\label{electric_current}
{\bf J}_{e}=\frac{iq\hbar}{2m}\left(\Psi\nabla\Psi^{*}-\Psi^{*}\nabla\Psi\right)-\frac{q^{2}}{mc}{\bf A}\Psi^{*}\Psi.
\end{equation}\\
However, due to the degeneracy of the system, working with the general solution is a challenging task. Nevertheless, we can make two considerations that are helpful in simplifying this situation. One consideration is regarding the structure of the constants involved in Eq. (\ref{Gen_sol}), and the other consideration is based on the assumption that the system temperature is lowered to its ground state.
Considering that when the quantization of the resistivity is observed, the longitudinal resistivity vanishes \cite{klitzing1980new,tsui1982two,willett1987observation,stormer1999fractional,stormer1999nobel,dean2008contrasting,von2017quantum}, it is equivalent to saying that the movement of the particle along the $y$-axis ceases. This condition implies that $c_{n,j,j'}=0$. On the other hand, the remaining constants can be selected in a way that we recover the unitary operators in Eq. (\ref{unitary1}) and Eq. (\ref{unitary3}), i.e.
\begin{equation}\label{}
\overline{c}_{n,j,j'}=\overline{c}_{n}\frac{1}{j!}\frac{\delta x^{j}}{(i\hbar)^{j}}\frac{1}{j'!}\frac{\delta t^{j'}}{(-i\hbar)^{j'}}.
\end{equation}\\
Then, since the system is expected to be in its ground state, i.e., $\overline{c}_{n}=0$ for all $n\neq 0$ and $\overline{c}_{0}=1$, the general solution can be written as 
\begin{equation}\label{ground_state}
\Psi=\overline\zeta_{0}(x-\delta x,y,t-\delta t).
\end{equation}
Defining $\Delta x=x-\delta x$, $\Delta t=t-\delta t$ and using Eq. (\ref{electric_current}) to calculate the electric current produced by this ground state, one can write down 
\begin{equation}\label{Je_electriccurrent_2d}
{\bf J}_{e}=\left(\frac{q^{2}}{\hbar}\frac{\hbar}{m\omega_{c}}{\cal E}{\hat i}-q\omega_{c}\left( \Delta x-\frac{q{\cal E}}{m\omega_c}\Delta t\right){\hat j}\right)|\Psi|^{2}.
\end{equation}\\
By definition, the current per unit of electric field, ${\bf J}_{e}/{\cal E}$, is equal to the conductivity. Also, note that by the way the system was set up, the coordinate along the $x$-axis represents the Hall conductivity, and the coordinate along the $y$-axis represents the longitudinal conductivity. Therefore, the Hall resistivity produced by this ground state is
\begin{equation}\label{}
\rho_{H}=\frac{\hbar}{q^{2}}\frac{m\omega_{c}}{\hbar}\frac{1}{|\Psi|^{2}},
\end{equation}
calculating the expected value of the above equality inside an area $A=\delta x\delta y$, gives the next result
\begin{equation}\label{QH_resistivity}
\braket{\rho_{H}}=\frac{\hbar}{q^{2}}\frac{m\omega_{c}}{\hbar}\delta x\delta y.
\end{equation}
As we mention in the previous section, in Heisenberg picture the Schr\"odinger equation is invariant under the unitary transform given by the operator Eq. (\ref{unitary2}), however, in the Schr\"odinger picture, the same transformation acting on the ground state Eq. (\ref{ground_state}) gives 
\begin{equation}\label{}
{\hat U}_{y}\Psi=\exp\left(-i\frac{m\omega_{c}}{\hbar}\delta x\delta y\right)\exp\left(i\frac{q{\cal E}}{\hbar}\delta t\delta y\right)\Psi.
\end{equation} 
Hence, the ground state is invariant under this unitary transformation only if the following conditions are satisfied 
\begin{equation}\label{fieldBQ}
\frac{m\omega_{c}}{\hbar}\delta x\delta y=2\pi l,\quad l\in\mathbb{N},
\end{equation} 
and
\begin{equation}\label{fieldEQ}
\frac{q{\cal E}}{\hbar}\delta t\delta y=2\pi k,\quad k\in\mathbb{N}.
\end{equation} 
Therefore, when this invariant condition is met, the Hall resistivity, Eq. (\ref{QH_resistivity}), is quantized in integer multiples of the von Klitzing's constant 
\begin{equation}\label{}
\braket{\rho_{H}}=\frac{h}{q^{2}}l,
\end{equation}
where $h=2\pi\hbar$. On the other hand, the longitudinal resistivity is given by the expression 
\begin{equation}\label{}
\rho_{L}=\frac{{\cal E}}{q\omega_{c}}\left( \Delta x-\frac{q{\cal E}}{m\omega_c}\Delta t\right)^{-1}\frac{1}{|\Psi|^{2}},
\end{equation}
it will vanish when the time is such that 
\begin{equation}\label{delT}
\Delta t<<\frac{m\omega_{c}}{q{\cal E}}\Delta x.
\end{equation}
The idea that the longitudinal resistivity vanishes after a given time interval can already be found in the literature as the relaxation time condition \cite[p.~498]{kittel}. However, there is a difference with this last expression, and that is that the time interval is bounded. 


\section*{Conclusions}
The conserved operators of the non-relativistic Hamiltonian with a constant electromagnetic field were analyzed, showing that they are all necessary to obtain a full description of the system; otherwise, the Lorentz force cannot be recovered. Simultaneously, the conserved operators in Eq. (\ref{px}) and Eq. (\ref{py}) were both used to find solutions to the Schrödinger equation, resulting in two basic wave functions: one for oscillations along the $y$-axis (which corresponds to Landau's solution) and a second one for oscillations along the $x$-axis. These two oscillations, each along their respective axis, can be thought of as projections of the classical circular movement of the particle. Continuing with the analysis, it was shown that the conserved operators defined a discrete degeneracy of the system, which was later used to find the general solution of the system as a linear combination of all the solutions.\\

\noindent Finally, a ground state was deduced from this general solution after a specific consideration of the constants $c_{n,j,j'}$ and $\overline {c}_{n,j,j'}$ involved in the linear combination. This state has the properties of having a quantized Hall resistivity in integer multiples of the von Klitzing's constant if it is invariant under the unitary transformation of the operator Eq. (\ref{unitary2}). Additionally, for the longitudinal resistivity, it was shown that it vanishes if the time interval condition in Eq. (\ref{delT}) is satisfied.

\newpage
\bibliographystyle{apsrev4-1}
\bibliography{bibliografia}

\end{document}